\documentclass[aip,amsmath,amssymb,prereprint,floatfix]{revtex4-1}
\usepackage{graphicx}
\usepackage{dcolumn}
\usepackage{bm}

\usepackage[utf8]{inputenc}
\usepackage[T1]{fontenc}
\usepackage{mathptmx}
\usepackage{etoolbox}
\usepackage{float}
\makeatletter
\let\newfloat\newfloat@ltx
\makeatother
\usepackage{algorithm}
\usepackage{algpseudocode}

\makeatletter
\def\@email#1#2{%
 \endgroup
 \patchcmd{\titleblock@produce}
  {\frontmatter@RRAPformat}
  {\frontmatter@RRAPformat{\produce@RRAP{*#1\href{mailto:#2}{#2}}}\frontmatter@RRAPformat}
  {}{}
}%
\makeatother

\begin{document}

\title{Simulation-Free Determination of Microstructure Representative Volume Element Size via Fisher Scores}
\author{Wei Liu}
\affiliation{ 
Department of Industrial Engineering and Management Sciences, Northwestern University, Evanston, Illinois 60208, USA
}%
\author{Satyajit Mojumder}%
\affiliation{ 
Theoretical and Applied Mechanics Program, Northwestern University, Evanston, Illinois 60208, USA
}%

\author{Wing Kam Liu}
\affiliation{ 
Department of Mechanical Engineering, Northwestern University, Evanston, Illinois 60208, USA
}%

\author{Wei Chen}
\affiliation{ 
Department of Mechanical Engineering, Northwestern University, Evanston, Illinois 60208, USA
}%
\author{Daniel W. Apley}
\email{apley@northwestern.edu.}
\affiliation{ 
Department of Industrial Engineering and Management Sciences, Northwestern University, Evanston, Illinois 60208, USA
}%


\begin{abstract}
A representative volume element (RVE) is a reasonably small unit of microstructure that can be simulated to obtain the same effective properties as the entire microstructure sample. Finite element (FE) simulation of RVEs, as opposed to much larger samples, saves computational expense, especially in multiscale modeling. Therefore, it is desirable to have a framework that determines RVE size prior to FE simulations. Existing methods select the RVE size based on when the FE-simulated properties of samples of increasing size converge with insignificant statistical variations, with the drawback that many samples must be simulated. We propose a simulation-free alternative that determines RVE size based only on a micrograph. The approach utilizes a machine learning model trained to implicitly characterize the stochastic nature of the input micrograph. The underlying rationale is to view RVE size as the smallest moving window size for which the stochastic nature of the microstructure within the window is stationary as the window moves across a large micrograph. For this purpose, we adapt a recently developed Fisher score-based framework for microstructure nonstationarity monitoring. Because the resulting RVE size is based solely on the micrograph and does not involve any FE simulation of specific properties, it constitutes an RVE for any property of interest that solely depends on the microstructure characteristics. Through numerical experiments of simple and complex microstructures, we validate our approach and show that our selected RVE sizes are consistent with when the chosen FE-simulated properties converge.
\end{abstract}

\maketitle

\section{\label{sec:level1}INTRODUCTION}
With advances in experimental characterization techniques (e.g., X-ray Computed Tomography (XCT), Scanning Electron Microscopy (SEM), Transmission Electron Microscopy (TEM)) over the last few decades, high-fidelity material microstructure images (aka micrographs) are increasingly available for understanding material systems. Analyses of the high-resolution micrographs can reveal important physics, e.g. related to failure and damage, of the materials. Finite element (FE) analysis is a popular method to approximately compute desired properties, such as stress/strain fields, over the sample domain by formulating a boundary value problem and solving a system of algebraic equations\cite{Liu22}. However, high-resolution finite element (FE) modeling of large microstructure samples is computationally prohibitive. Instead of simulating the entire microstructure sample, it typically suffices to simulate only a representative volume element (RVE) of the material. Loosely speaking, an RVE is a sample that is reasonably small but still large enough to be representative of the entire microstructure, in the sense that its simulated properties are the same as those of a large sample. The fine-scale simulation of an RVE is also essential in approaches regarding multiscale modeling\cite{Wagner21, MCVEIGH20065053}. For example, RVE-size cells are assigned to macrostructural locations before the system is simulated at the macro-level\cite{GHOSH20012335}. Or as in the self-consistent clustering analysis method\cite{LIU2016319}, offline, linear elastic simulations of an RVE are used for both clustering on computational cells and online modeling. RVE concepts have also been combined with multi-fidelity simulation to reduce computational expense in assessing aleatoric uncertainty due to microstructure variability by simulating multiple smaller-than RVE samples at multiple fidelity levels\cite{Anh231, Anh232}. Therefore, a framework to determine the appropriate RVE size for a given microstructure prior to conducting any FE simulation can substantially improve simulation efficiency, as it avoids simulating unnecessarily large samples. 

By definition, an RVE is just large enough that local statistical variations of microstructure properties across the RVE average out, whereas such variations across smaller samples do not. Definitions of an RVE in the literature vary in specifics but coincide in their essence. Some literature defines an RVE via constitutive laws. Drungan and Willis\cite{DRUGAN1996497} defined an RVE as an element "for which the usual spatially constant (overall modulus) macroscopic constitutive representation is a sufficiently accurate model to represent the mean constitutive response." As the mean constitutive response is superimposed with large fluctuations related to local microstructural detail, they took the RVE size to be just large enough that the corrections to the local terms in the explicit nonlocal constitutive equations are negligible. They derived an estimator of the minimum RVE size and showed that it is twice the reinforcement diameter for two-phase composites with isotropic and statistically uniform distribution of arbitrarily shaped, arbitrarily anisotropic phases. Alternatively, Ostoja-Starzewski\cite{OSTOJASTARZEWSKI2006112} employed the Hill condition, which requires a valid RVE to satisfy
\begin{align}\label{eqn:Hill}
    \langle \bm{\sigma} : \bm{\bm{\epsilon}}\rangle = \langle\bm{\sigma}\rangle : \langle\bm{\bm{\epsilon}}\rangle
\end{align}
where $\langle\cdot\rangle$ is the mean operator, $\bm{\sigma}$ the stress tensor, and $\bm{\epsilon}$ the strain tensor. The above definitions implicitly require statistical stability of the mechanical properties of RVEs. Probability and statistics concepts were more explicitly incorporated to characterize an RVE in the following works. Sab and Nedjar\cite{SAB2005187} defined the RVE size as when the homogenized elasticity tensor converges up to a given relative error and estimated the minimum RVE size based on the two-point correlation functions. Kanit \textit{et al.}\cite{Kanit2003} defined RVE as a volume of heterogeneous material sufficiently large to be statistically representative of the composite. Similar definitions are commonly seen in recent literature, though different works use different definitions of "statistically representative". To numerically determine RVE size for a given microstructure, various statistical analyses are applied to things like morphology features, FE-simulated properties\cite{GREENE2013271}, or measurements of Digital image correlation (DIC) \cite{KOOHBOR201759}. Kanit \textit{et al.}\cite{Kanit2003} proposed a statistical framework in which the general idea is to select the RVE size by simulating properties (or calculating morphology features) in samples of different sizes and choosing the size at which the simulated properties converge in terms of their statistical fluctuations being sufficiently small. As most relevant works are adapted from the framework in Kanit, \textit{et al.}\cite{Kanit2003} and focus on the convergence of FE-simulated properties, we will refer to this class of approaches as the simulation method, because it requires FE simulation of many samples of increasing size. 

The simulation method is general and has been applied to different material systems such as random quasi-brittle composites \cite{pelissou2009}, polymers \cite{MIRKHALAF201630, HU201855}, fiber bundles \cite{STAPLETON2016170}, and others\cite{Grimal, ERASLAN2022111070, KRISHNAMOHAN2020112633}.  Various criteria for assessing convergence of simulated properties in terms of their statistical fluctuations have been considered, e.g., confidence intervals \cite{Kanit2003, HU201855, SALMI2012230} and Chi-square hypothesis tests \cite{gitman2007}. Although the simulation method is straightforward and widely accepted, its major disadvantage is that it requires many FE simulations, since many different sample sizes must be simulated with each sample size simulated multiple times (to assess the variation in properties for different samples of the same size), typically including sizes larger than an RVE, which are computationally expensive. In some settings, this defeats the purpose of RVE size determination, if the goal is to avoid excessive computational expense when simulating properties. Moreover, it inextricably links RVE size to the simulated property of interest\cite{REN2002881} and, for the same material, can result in two different RVE sizes for two different material properties. For example, Kanit \textit{et al.}\cite{Kanit2003} showed that when simulating thermal and elastic properties, the corresponding RVE sizes differ. Savvas, Stefanou, and Papadrakakis\cite{SAVVAS2016340} found that the apparent shear modulus converges much faster than the apparent axial stiffness, and the two properties have different RVE sizes. Trias \textit{et al.}\cite{TRIAS20063471} conducted a comprehensive comparison study considering different criteria and properties and concluded a similar inconsistency. 

To avoid the disadvantages of the simulation method, we propose a new approach that aims to determine RVE size in a manner that does not require any prior FE simulation and applies to any property of interest, in the sense that any simulated property of the selected RVE will be representative of a large sample. This is particularly useful when multiple properties are of interest. To determine RVE size without FE simulation, our algorithm only requires a micrograph whose size is larger than an RVE (size requirements are discussed in Sec. \ref{sec:plot}). The micrograph can be an original grayscale image (such as experimentally collected micrographs), a color image, or a discretized version whose pixels have been converted (through any standard preprocessing analysis) to discrete phase values for a multiphase composite image.

We note that Sab and Nedjar\cite{SAB2005187} approach mentioned earlier also determines RVE size without using FE simulation. However, their approach was restricted to linear elastic materials, since they incorporate the physics of homogenized elasticity tensors into their RVE size derivations. In contrast, our approach is intended to be more generally applicable to any property of interest, governed by either nonlinear or linear physics, providing that the property is determined solely by the morphology (micrograph) of the material. 

To accomplish this, our approach is based on the notion of Fisher score vectors for an appropriate machine learning (ML) model fitted to the micrograph. Specifically, we fit an ML model to predict the value for each pixel in the micrograph as a function of its surrounding pixel values. The Fisher score vector, which we refer to as the score vector for short, is the derivative of the log-likelihood for that pixel with respect to (w.r.t.) the ML model parameters, and is computed for each pixel. The fundamental idea is that the ML model provides an implicit representation, or fingerprint, of the stochastic nature of the microstructure, and the score vector for each pixel provides an indication of how much the local stochastic nature at each pixel differs from the aggregate stochastic nature across the entire micrograph. Prior work\cite{ZHANG2021116818} used score vector concepts to monitor the nonstationarity of microstructure behavior and showed that if the mean score vector over a local window (representing a subregion) of the micrograph is nonzero, the stochastic nature of the microstructure in that local window differs from the aggregate behavior over the entire micrograph, which is an indication of nonstationarity. 

For our purpose of RVE size determination, we use the score vectors in a different, but related manner. We cast the problem of determining the RVE size as investigating the nonstationarity of moving-window samples of varying window sizes, as the window moves across the micrograph. The rationale is that, for a specified window size, if the microstructure sample represented by the moving window is stationary as the window moves across the micrograph, i.e., if the stochastic nature of the microstructure within the moving window is similar wherever the window is positioned across the entire micrograph, then this window size is at least as large as the RVE size. In this sense, that sized window constitutes a "representative" sample of the microstructure, and any computed material property for the window that depends on the aggregate behavior of the microstructure over the window should be the same as for the entire micrograph. We assess the stationarity of the moving window samples via the score-based framework, as described later. 

Stationarity is not a well-defined black-and-white measure whereby the behavior within a particular window becomes completely stationary when the window size increases beyond a threshold. Rather, virtually any microstructure monotonically becomes less and less nonstationary as the window size increases and, for sufficiently large enough window sizes, is essentially stationary. Consistent with this, our score-based approach involves plotting a measure of nonstationarity versus window size as window size grows (e.g., Figs. \ref{fig:fig3}, \ref{fig:fig5}, \ref{fig:fig7}), and then identifying the window size beyond which nonstationarity further decreases only mildly, in some sense. Our experimental results indicate that this plot agrees quite closely with the convergence of FE-simulated properties as window size grows. Technically, the RVE size determined by our method should be viewed as an upper bound on the RVE size, since smaller windows may have the same value for a property wherever the window is positioned, even if the stochastic nature of the microstructure differs (which might happen for structure-insensitive properties that do not significantly depend on specific arrangement of particles\cite{stroeven2004}). Our experimental results indicate that it provides a relatively tight upper bound (see Sec. \ref{sec:results}). Because our approach operates on only micrographs without the need for FE simulation, our RVE size determination is based only on the underlying microstructure instead of being tied to a specific simulated property.
    
The remainder of the paper is organized as follows. In Sec. \ref{sec:score-based}, we provide background on the score-based framework and its prior use for nonstationarity monitoring. In Sec. \ref{sec: method}, we detail our approach to adapt the score-based concepts to determine RVE size. In Sec. \ref{sec:results}, we present experimental results for 2D micrographs of two-phase composite microstructures to illustrate our method. Sec. \ref{sec: conclusion} concludes the paper.

\section{Background on the score-based framework for microstructure nonstationarity monitoring}\label{sec:score-based}
This section overviews prior work that developed the score-based framework for monitoring nonstationarity of microstructures \cite{ZHANG2021116818, Kungang1}. Let the scaler variable $Y$ denote the value of an individual pixel in the micrograph, and let the vector variable $\bm{X}$ denote the values for some appropriate set of neighboring pixels surrounding it. Let $y$ and $\bm{x}$ denote realizations of $Y,\bm{X}$, respectively, i.e., a particular pixel and its surrounding neighbors. Given an arbitrary micrograph, the conditional distribution $P(Y|\bm{X})$ of $Y$ given $\bm{X}$, which models the predictive distribution of pixel $Y$ given its neighboring pixels $\bm{X}$, implicitly represents the joint distribution of all pixels in the micrograph \cite{BOSTANABAD201689, BOSTANABAD20181}, which is a full characterization of the stochastic nature of the microstructure. We can, therefore, analyze microstructure nonstationarity by analyzing changes in $P(Y|\bm{X})$ as the position of pixel $Y$ moves across the micrograph.

$P(Y|\bm{X})$ can be estimated or approximated using an appropriate ML model. For binary micrographs of two-phase composite materials (e.g., matrix with particles), $P(Y|\bm{X})$ is represented via a binary classification ML model. For more general grayscale images, one can approximate $P(Y|\bm{X})$ as a normal distribution (for traceability purposes) with mean modeled as a function of $\bm{x}$\cite{BOSTANABAD201689}, i.e., $Y|\bm{X}=\bm{x}\sim\mathcal{N}(f(\bm{x},\hat{\bm{\theta}}), \sigma^2)$, where $f(\bm{x},\hat{\bm{\theta}})$ is an ML regression model for predicting $Y$ as a function of $\bm{X}$, and $\hat{\bm{\theta}}$ are the trained parameters of the model (e.g., the weights and bias coefficients of a neural network). In this paper, we focus on micrographs for two-phase composite materials and use a binary classification ML model $f(\bm{x},\hat{\bm{\theta}})$ to approximate $P(Y|\bm{X}=\bm{x})$ directly. 

\begin{figure*}[ht!]
    \centering
    \includegraphics[width=0.8\textwidth]{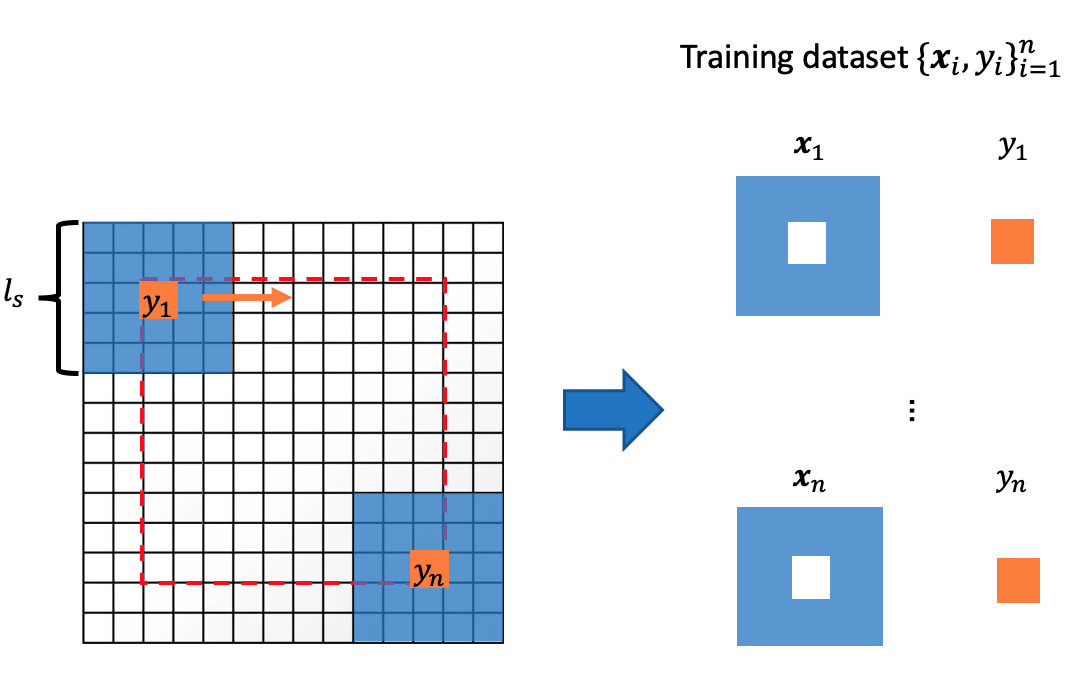}
    \caption{Illustration of how the training dataset is constructed from the micrograph. Each pixel $i$ in the red dashed box is associated with one data point $(\bm{x}_i, y_i )$. The parameter $l_s$ represents the neighborhood size and determines the dimension ($l_s^2-1$) of $\bm{x}_i$.}
    \label{fig:fig0}
\end{figure*}

Using this ML-based framework, Zhang, Apley, and Chen\cite{ZHANG2021116818} converted the problem of analyzing nonstationarity in $P(Y|\bm{X})$ to analyzing nonstationarity in the ML model parameters $\hat{\bm{\theta}}$. They developed a score-based approach in which one fits a single ML model $f(\bm{x};\hat{\bm{\theta}})$ to a training dataset $\{\bm{x}_i,y_i\}_{i=1}^n$ (see Fig. \ref{fig:fig0}) constructed from the entire micrograph, where $y_i$ is the value of the pixel with location indexed by $i$, $\bm{x}_i$ is the corresponding input vector composed of neighboring pixels (the neighborhood size parameter $l_s$ can be chosen via cross-validation when fitting the ML model, or via other means), and $n$ is the total number of pixels in the micrograph (excluding pixels on the boundary that do not have a full neighborhood of surrounding pixels). The score-based framework then computes the score vector at each pixel $i$, which is defined as the derivative of the log-likelihood of $y_i$ w.r.t. $\hat{\bm{\theta}}$:
\begin{align}\label{eqn:score}
    \mathbf{s}(\hat{\bm{\theta}};y_i, \mathbf{x}_i)=\nabla_{\hat{\bm{\theta}}}\log P(Y=y_i|\mathbf{X}=\mathbf{x}_i;\hat{\bm{\theta}})
\end{align}
where $\nabla_{\hat{\bm{\theta}}}$ denotes the gradient operator w.r.t. $\hat{\bm{\theta}}$. Note that the log-likelihood in Eq. (\ref{eqn:score}), when summed over all $n$ training observations, is the same log-likelihood function that is maximized when estimating $\bm{\theta}$. Note also that if some form of regularization is used when fitting the model, which is common, then $P(Y|\bm{X})$ in Eq. (\ref{eqn:score}) should be replaced by its regularized version, so that the gradient of the model fitting objective function coincides with the sum of the score vectors in Eq. (\ref{eqn:score}) across all $n$ training observations (see Ref. \onlinecite{Kungang1}).

Fundamental statistical theory implies that if the predictive relationship $P(Y|\bm{X})$ is stationary across the micrograph, the score vector is zero-mean at each pixel location $i$. Conversely, if $P(Y_i|\bm{X_i})$ is nonstationary, then the mean of the score vector will differ from the zero vector, denoted as $\bm{0}$, at the locations at which $P(Y_i|\bm{X_i})$ changes. The score-based framework\cite{ZHANG2021116818} uses this fact to monitor for nonstationarity by computing the average score (which serves as a local estimate of the mean score vector) within some appropriately small moving window centered around each pixel in the micrograph, and then using statistical tests to determine at which locations the score vector mean differs from $\bm{0}$. One then concludes that the stochastic nature of the microstructure is different at those locations. Similar concepts were used to segment nonstationary micrographs into different regions that each has distinct microstructure characteristics\cite{ZHANG2021116818}. Note that this framework can monitor for concept drift in more general ML modeling situations. In the ML literature, concept drift refers to changes or drift over time/space in the predictive relationship that governs the data and that is represented by a fitted ML model. Refer to Refs. \onlinecite{ZHANG2021116818, Kungang1} for further details on concept drift and various applications of the score-bared framework.

\section{Methodology}\label{sec: method}
\begin{figure*}[ht!]
    \centering
    \includegraphics[width=\textwidth]{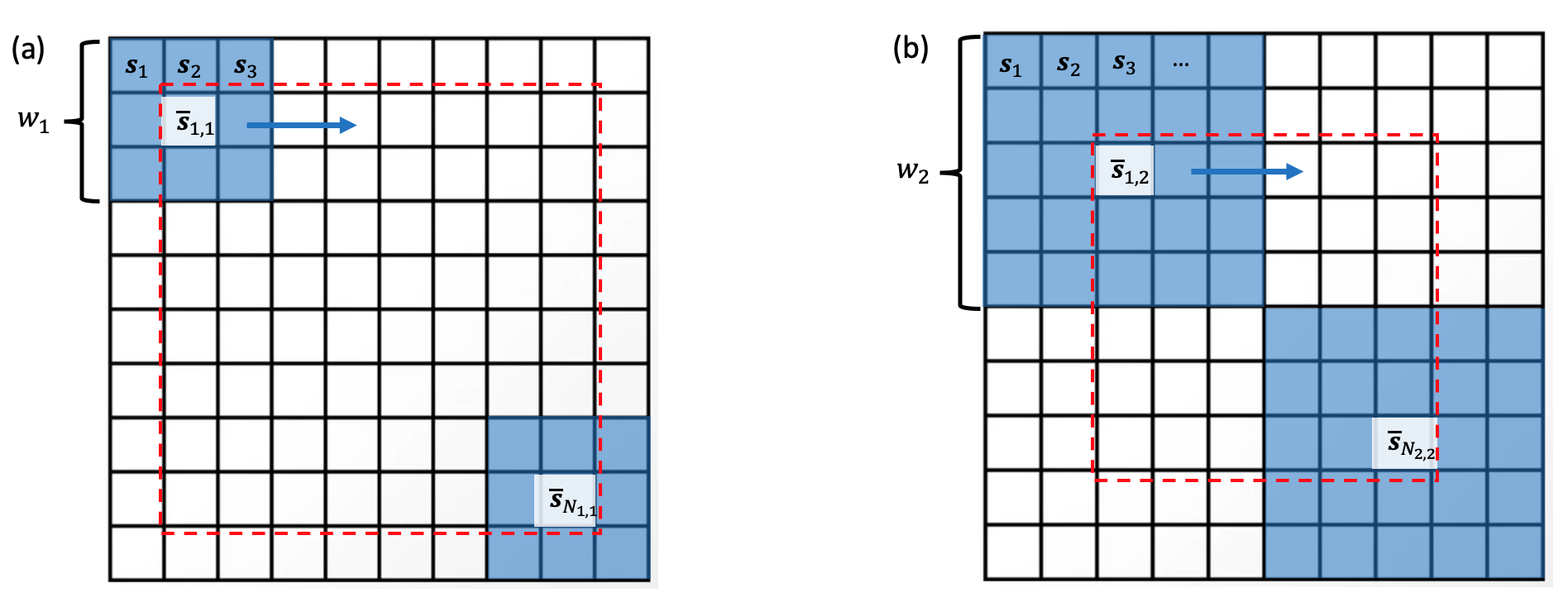}
    \caption{Comparison of two candidate RVE sizes, $w_1=3$ pixels and $w_2=5$ pixels, for moving windows (depicted as blue shaded squares at two locations in each image). The interiors of the red dashed boxes represent the pixels around which the moving window can be centered without extending outside the micrograph, the total number of which is $N_k$. Each $\bar{\bm{s}}_{i,k}$ denotes the average score vector across all pixels in the shaded moving window of size $w_k$, centered at pixel $i$.}
    \label{fig:fig1}
\end{figure*}

Algorithm \ref{alg: algo} presents pseudocode for our proposed algorithm for RVE size determination, the details of which are elaborated in the following subsections. The main ideas behind the algorithm are as follows. We first fit a supervised learning model to the entire micrograph and compute its score vector at each pixel in the micrograph (Sec. \ref{sec:model}). To determine RVE size, consider an arbitrary moving window size $w$ as depicted in Fig. \ref{fig:fig1} (unless otherwise noted, window size will refer to the length $w$ of a side of the square window) as a candidate RVE size. If this window is at least as large as an RVE, then by the arguments in Sec. \ref{sec:score-based}, the average score vector across the window (described in Sec. \ref{sec:window}) should be close to $\bm{0}$, and this should hold wherever the moving window is positioned across the entire micrograph. On the other hand, if this window size is smaller than an RVE, then the average score vector across the window should differ substantially from $\bm{0}$ at many moving window positions across the micrograph. 

Consequently, our approach is to compute some measure (denoted by $\overline{D}_k$, described in Sec. \ref{sec:t_2}) of the extent to which the average score vector across the moving window of $w=w_k$ differs from $\bm{0}$ as the window moves across the entire micrograph. We repeat these computations for each candidate RVE size $w_k \in \{w_1, w_2, ..., w_K\}$, where $w_1$ represents the smallest candidate size and $w_K$ the largest. We then construct a plot of $\overline{D}_k$ vs. $w_k$ (Sec. \ref{sec:plot}) and use this plot to decide the smallest $w_k$ for which the average score vector is sufficiently close to $\bm{0}$, which we take to be the RVE size. Once the RVE size is determined, users can conduct a FE simulation of a single sample of that size to determine the properties of the microstructure.

\begin{algorithm}
\caption{Pseudocode for the RVE Size Determination Algorithm}\label{alg: algo}
\begin{algorithmic}
\Require A micrograph that is large enough to select an RVE from.
\Require A list of $K$ candidate RVE sizes $\{w_1, w_2, ... w_K\}$. \Comment{Fig. \ref{fig:fig1}}
\State Fit a parametric ML model $f(\bm{x};\hat{\bm{\theta}})$ to the training dataset $\{\bm{x}_i, y_i\}_{i=1}^n$ constructed from the input micrograph;
\State Calculate the score vectors $\{\bm{s}_1,...,\bm{s}_n\}$ at each pixel; \Comment{Sec. \ref{sec:model}}
\For{$k$ in $\{1, 2, ..., K\}$}
\State Compute $\{\bar{\bm{s}}_{i,k}\}_{i=1}^{N_k}$, where $\bar{\bm{s}}_{i,j}$ is the sample average of the score vectors across the pixels within the size-$w_k$ window centered at pixel $i$, and $N_k$ denotes the total number of positions within the micrograph for the size-$w_k$ window; \Comment{Sec. \ref{sec:window}}
\State Compute $D_{i,k}$, $i\in\{1,...,N_k\}$; \Comment{Sec. \ref{sec:t_2}}
\State Compute $\overline{D}_k=\frac{1}{N_k}\sum_{i=1}^{N_k} D_{i,k}$; \Comment{Sec. \ref{sec:t_2}}
\EndFor
\State Plot $\overline{D}_{k}$ vs. $w_k$ and select the RVE size based on the elbow in the plot. \Comment{Sec. \ref{sec:plot}}
\end{algorithmic}
\end{algorithm}

\subsection{ML model fitting and computation of the score vector $\bm{s}_i$}\label{sec:model} 
Given the micrograph, we fit a supervised ML model $f(\bm{x};\hat{\bm{\theta}})$ to the corresponding dataset (Fig. \ref{fig:fig0}). Here $\hat{\bm{\theta}}$ is the maximum-likelihood estimator of the ML model parameters. As discussed earlier, $f(\bm{x}_i;\hat{\bm{\theta}})$ predicts the value $y_i$ of each micrograph pixel as a function of its neighboring pixels $\bm{x}_i$ and approximates the predictive distribution $P(y_i|\bm{x}_i;\bm{\theta})$. Then, for each pixel $i$ in the micrograph (excluding the boundary pixels for which there is no full neighborhood, as depicted in Fig. \ref{fig:fig0}), its score vector $\bm{s}_i$ is calculated via Eq. (\ref{eqn:score}). The ML model $f(\bm{x}_i;\hat{\bm{\theta}})$, together with $\bm{s}_i$, serve as a fingerprint of the local stochastic nature around pixel $i$.

The score-based framework, and thus our method, can only be applied to parametric ML models. For the micrographs considered in the experiments in Sec. \ref{sec:results}, we used logistic regression and neural network models. These two models are natural choices as our samples are two-phase binary images (particles dispersed in the matrix) requiring binary classifiers. When the micrographs have discrete pixel values but are more complex (e.g., with three phases: voids, particles, and matrix), one can use a multi-class ML classifier such as multinomial logistic regression or a multi-class neural network classifier\cite{ESL}. For grayscale images with a large number of pixel values, one should use a parametric regression ML model such as a neural network of linear regression model. For neural networks, we have found that computing the score vector for only the parameters in the last layer (which should be fully connected) of the neural network, i.e., the gradient of the log-likelihood w.r.t. only the parameters of the last layer, is effective for monitoring stationarity and is far more computationally efficient and stable\cite{Kungang1}, and hence we recommend this approach and have used it in our examples.

\subsection{Computation of the average score vector $\bar{\bm{s}}_{i,w}$}\label{sec:window}
For a candidate RVE window size $w_k$ (i.e., a square window having $n_k=w_k^2$ pixels), denote the average score vector across the window centered at pixel $i$ by $\bar{\bm{s}}_{i, k}$. Fig. \ref{fig:fig1} illustrates this for two candidate RVE sizes, $w_1=3$ ($n_1=9$) pixels and $w_2=5$ ($n_2=25$) pixels. For each window size, moving windows are shown at two locations within the micrograph. Note that for candidate window size $w_k$, the total number of micrograph pixels around which the window can be centered without extending outside the micrograph, which we denote by $N_k$, decreases as $w_k$ increases. For example, in Fig. \ref{fig:fig1}, for $w_1=3$, $N_1=64$, and for $w_2=5$, $N_2=36$. For each $k \in \{1, 2, ..., K\}$ (i.e., for each candidate RVE size $w_1, w_2, ..., w_K$), we compute $\{\bar{\bm{s}}_{i,k}: i = 1, 2, ..., N_k\}$.

\subsection{Measuring the extent to which $\{\bar{\bm{s}}_{i,k}: i=1,2,...,N_k\}$ differs from $\bm{0}$}\label{sec:t_2}
As mentioned earlier, for candidate RVE size $w_k$, we must measure the extent to which the average score vectors $\{\bar{\bm{s}}_{i,k}: i = 1, 2, ..., N_k\}$ differ from $\bm{0}$ across the micrograph. If they differ substantially, we conclude that $w_k$ is too small to be the RVE size. We modify this slightly by considering the extent to which $\{\bar{\bm{s}}_{i,k} - \bar{\bar{\bm{s}}}: i = 1, 2, ..., N_k\}$ differ from $\bm{0}$ across the micrograph, where $\bar{\bar{\bm{s}}}$ is the sample mean vector of $\{\bm{s}_1,...,\bm{s}_n\}$. Technically, if the optimization in the maximum likelihood (or regularized maximum likelihood) fitting of the ML model is done to completion with no constraints on the parameters, then $\bar{\bar{\bm{s}}}$ is exactly $\bm{0}$ barring numerical precision errors, since $\bar{\bar{\bm{s}}}$ is the derivative of the log-likelihood function. However, if one uses stochastic gradient descent with early stopping to fit the model and/or there are constraints on the parameters in the optimization, then $\bar{\bar{\bm{s}}}$ will typically not be $\bm{0}$ exactly. Considering the extent to which $\{\bar{\bm{s}}_{i,k} - \bar{\bar{\bm{s}}}: i = 1, 2, ..., N_k\}$ differ from $\bm{0}$ takes this into account.

One might consider using more formal statistical tests for this, e.g., by invoking the central limit theorem to approximate each $\bar{\bm{s}}_{i,k}$ as multivariate normal and computing some form of Hotelling $T^2$ statistic, e.g., $T^2_{i,k}=(\bar{\bm{s}}_{i,k} - \bar{\bar{\bm{s}}})^T\widehat{\bm{\Sigma}}_k^{-1}(\bar{\bm{s}}_{i,k} - \bar{\bar{\bm{s}}})$, where $\widehat{\bm{\Sigma}}_k$ denotes an estimate of the covariance matrix of $\bar{\bm{s}}_{i,k} - \bar{\bar{\bm{s}}}$. However, computing and inverting sample estimators of the covariance matrix is complicated by (i) the fact that the score vectors $\bm{s}_i$ at pixels $i$ located near each other are highly correlated and (ii) inversion of sample covariance matrices is notoriously badly behaved numerically. 

Consequently, we use a simpler and numerically better behaved measure of the extent to which $\{\bar{\bm{s}}_{i,k} - \bar{\bar{\bm{s}}}: i = 1, 2, ..., N_k\}$ differ from $\bm{0}$. Namely, we consider
\begin{align}\label{eqn:t_2 average}
    \overline{D}_k = \frac{1}{N_k}\sum_{i=1}^{N_k} D_{i,k},
\end{align}
where
\begin{align}\label{eqn:t_2}
    D_{i,k} = (\bar{\bm{s}}_{i,k} - \bar{\bar{\bm{s}}})^T\bm{A}^{-1}(\bar{\bm{s}}_{i,k} - \bar{\bar{\bm{s}}})
\end{align}
for some appropriate positive-definite matrix $\bm{A}$. Note that a small $\overline{D}_k$ indicates that there is relatively little variation of $\{\bar{\bm{s}}_{i,k}: i = 1, 2, ..., N_k\}$ about $\bar{\bar{\bm{s}}}$ (which implies that $w_k$ is large enough to be the RVE size), whereas a large $\overline{D}_k$ indicates that there is relatively large variation of $\{\bar{\bm{s}}_{i,k}: i = 1, 2, ..., N_k\}$ about $\bar{\bar{\bm{s}}}$ (which implies that $w_k$ is too small to be the RVE size).

We consider two choices for $\bm{A}$. One choice is $\bm{A} = \widehat{\bm{\Sigma}}_s$ where 
\begin{align*}
    \widehat{\bm{\Sigma}}_s = \frac{1}{n}\sum_{i=1}^n (\bm{s}_i - \bar{\bar{\bm{s}}})^T(\bm{s}_i - \bar{\bar{\bm{s}}})
\end{align*}
is the sample covariance matrix of $\{\bm{s}_1, ..., \bm{s}_n\}$. Another choice is its diagonalized version $\bm{A} = diag(\widehat{\bm{\Sigma}}_s)$. For high-dimensional $\hat{\bm{\theta}}$, the inversion of $diag(\widehat{\bm{\Sigma}}_s)$ is numerically much better behaved than the inversion of $\widehat{\bm{\Sigma}}_s$.  We consider both choices of $\bm{A}$ in the experimental results in Sec. \ref{sec:results}, and for these examples, both performed comparably. 

\subsection{A plot of $\overline{D}_k$ vs. $w_k$ to determine RVE size}\label{sec:plot}
It would be difficult to quantitatively select a threshold for $\overline{D}_k$, below which we can conclude $w_k$ is at least RVE size and above which we can conclude $w_k$ is smaller than RVE size. Instead, we have found that looking for the elbow in a plot of $\overline{D}_k$ vs. $w_k$ provides a robust and effective means of determining RVE size. In particular, we choose the RVE size to be the smallest $w_k$ that falls to the right of the elbow in the plot of $\overline{D}_k$ vs. $w_k$ (see Sec. \ref{sec:results} for examples), i.e., the smallest $w_k$ that falls into the region over which the plot $\overline{D}_k$ vs. $w_k$ has stabilized and remains relatively constant. This is analogous to the use of a scree plot in PCA, for graphically determining the number of dominant eigenvalues\cite{ESL}. Graphical methods like this for selecting parameters that are difficult to select via quantitative thresholds are commonplace in statistical analyses. Note that in PCA, instead of graphically identifying the elbow in a scree plot of the eigenvalues, one sometimes specifies a threshold on the ratio representing the cumulative sum of eigenvalues divided by the sum of all eigenvalues, which represents the proportion of total variance accounted for by the selected number of eigenvalues\cite{Peter2015}. The analogous quantity in our situation (area under the curve up to the elbow, divided by the total area under the curve) would not be meaningful, since it would depend on the choice of largest window size (e.g., $800$, $1200$, and $1000$ in Figs. \ref{fig:fig3}, \ref{fig:fig5}, \ref{fig:fig7}, respectively) and would have no physical meaning like the proportion-of-variance ratio in PCA. Fortunately, the experimental results in Sec. \ref{sec:results} indicate that our graphical approach for selecting RVE size coincides quite closely with when the FE-simulated properties converge.

Note that the only input to our algorithm is a sufficiently large micrograph. The micrograph must be large enough that a moving window with size $w_k$ equal to RVE size can be positioned at sufficiently many locations for the statistics $D_{i,k}$ and $\overline{D}_k$ to be meaningful. In our experiments in Sec. \ref{sec:results}, we used input micrographs that are $5\sim 8$ times the estimated RVE size. 

\section{Results and discussion}\label{sec:results}
We investigate and validate our algorithm on 2D micrographs of two-phase composite microstructures with varying inclusion sizes, shapes, and spatial distributions. In all the experiments, we use $l_s=21$ pixels to generate the training dataset. We consider two ML models -- logistic regression and a simple neural network with ten nodes in the hidden layer -- and use cross-validation (CV) to tune the neural network. For the neural network, we compute score vectors w.r.t. only the parameters in the last layer as discussed in Sec. \ref{sec:model}. Although our method does not use any FE simulation\cite{MCVEIGH20083268, YIN20083516} to determine the RVE size (which is one of its advantages), we do use FE simulation results to verify that our RVE choice coincides with when the FE results converge. 

\subsection{Microstructure with round-shaped particles}\label{sec:regular}
\begin{figure*}[ht!]
    \centering
    \includegraphics[width=\textwidth]{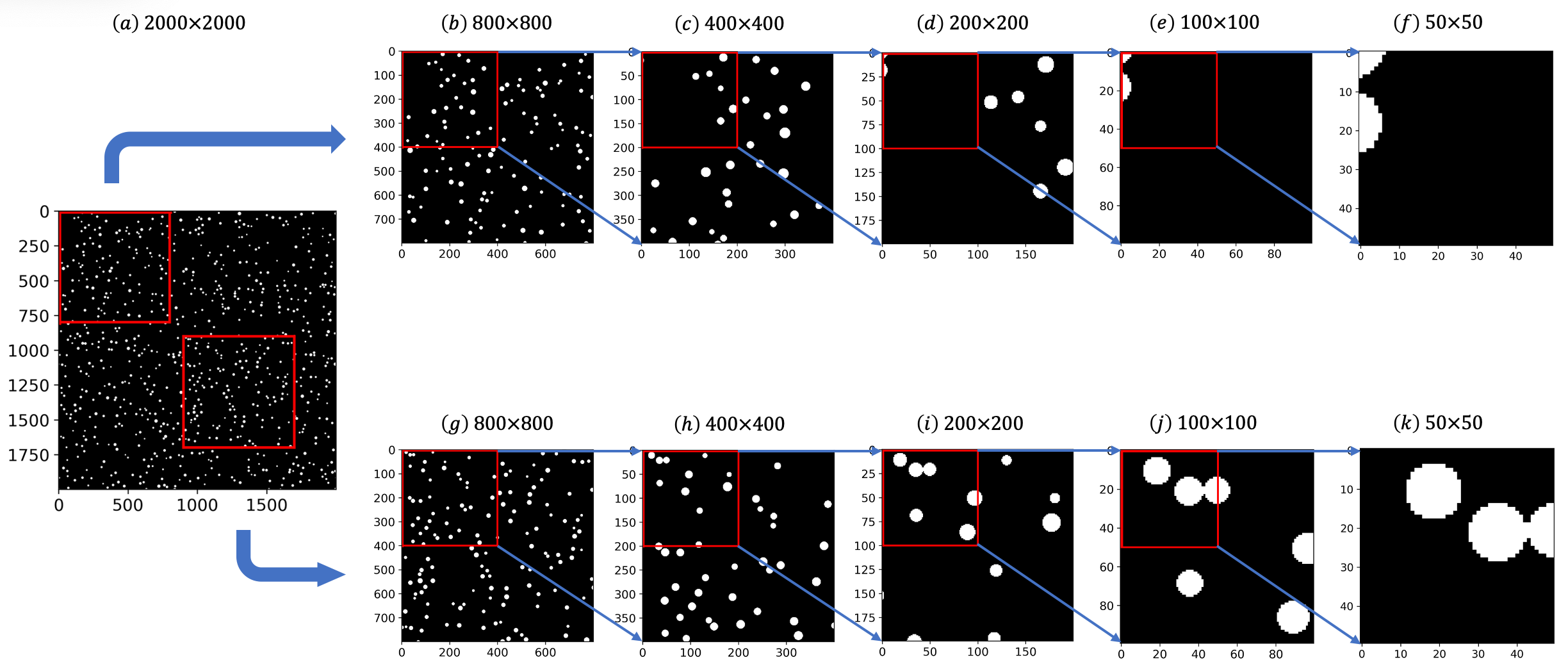}
    \caption{(a) A $2000 \times 2000$ pixel \text{micrograph} that represents a $2\mu m\times 2\mu m$ PMMA/SiO$_2$ microstructure with $3\%$ vf. (b - k) Magnified subregions of (a) for various window sizes.}
    \label{fig:fig2}
\end{figure*}

Fig. \ref{fig:fig2}(a) shows a micrograph (of size $2000\times 2000$ pixels) with round-shaped particles, which is representative of polymethyl methacrylate with silica particles (PMMA/SiO$_2$). The microstructure has $3\%$ volume fraction (vf) and $2 \mu m \times 2 \mu m$ physical size. Figs. \ref{fig:fig2}(b - k) show sequences of increasingly magnified subregions of the original micrograph, which are analogous to windows of decreasing size $w$. For the small sample size in Figs. \ref{fig:fig2}(f, k), the within-window stochastic nature would clearly not be stationary as the window moves across the micrograph (e.g., some window positions would contain no particles at all), which means this window would be smaller than an RVE. On the other hand, for the larger window size as in Figs. \ref{fig:fig2}(b, g), the within-window stochastic nature appears similar for different window positions, in which case this window would be at least as large as an RVE. As will be seen shortly, our method and the FE results both indicate that RVE size is somewhere in between these two window sizes.

\begin{figure*}[ht!]
    \centering
    \includegraphics[width=\textwidth]{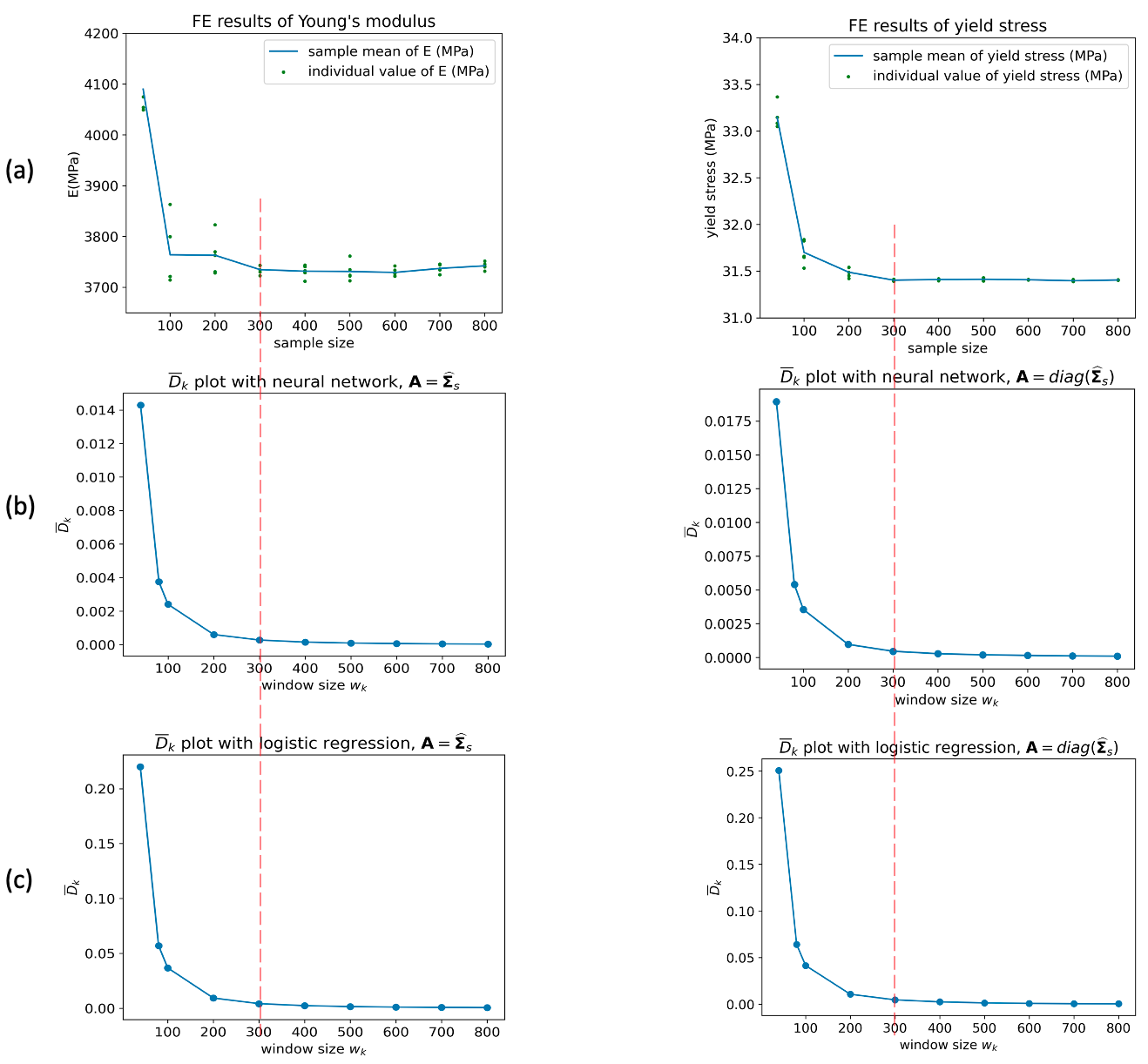}
    \caption{(a) FE-simulated Young's modulus and yield stress vs. sample size, with five samples simulated for each sample size. Green dots are the simulated properties of individual samples, and the blue line is the mean of the five individual sample results. (b - c) Plots of $\overline{D}_k^2$ vs. $w_k$ for a neural network ML model (b) and logistic regression ML model (c). Left column used $\bm{A} = \widehat{\bm{\Sigma}}_s$ and right column used $\bm{A}= diag(\widehat{\bm{\Sigma}}_s)$. The red vertical lines indicate the ground truth RVE size based on the FE results in (a).}
    \label{fig:fig3}
\end{figure*}

To provide the "ground truth" RVE size for comparison with the RVE size selected by our approach, Fig. \ref{fig:fig3}(a) shows FE results of Young's modulus and yield stress \cite{MOULINEC199869} of square samples of varying side length between $50$ to $800$ pixels. Based on the FE results in Fig. \ref{fig:fig3}(a), we somewhat subjectively determine the ground truth RVE size to be approximately $300$ pixels (after which the FE results appear to have converged), at which we have added a vertical dashed line in Fig. \ref{fig:fig3}. To determine RVE size using our approach, we used Algorithm \ref{alg: algo} with Fig. \ref{fig:fig2}(a) as the input micrograph and considered window size $w$ between $40$ and $800$ pixels. The resulting plots of $\overline{D}_k$ vs. $w_k$ for the neural network (CV accuracy $99.78$\%) and logistic regression (CV accuracy $99.91$\%) ML models are shown in Figs. \ref{fig:fig3}(b) and \ref{fig:fig3}(c), respectively. The left and right columns show results using $\bm{A} = \widehat{\bm{\Sigma}}_s$ and $\bm{A}= diag(\widehat{\bm{\Sigma}}_s)$, respectively. The above CV accuracies (as well as the CV accuracies stated for the later examples) were calculated as the balanced accuracy\cite{grandini2020metrics}, i.e., the average of the two CV correct classification accuracies for the two response classes (matrix vs. particle), since the training dataset is imbalanced ($3\%$ vf). 

Using our approach to select RVE size based on either of Figs. \ref{fig:fig3}(b, c), one would select approximately $w = 300$ pixels as the smallest $w$ that falls to the right of the elbow in the plot, i.e., the smallest $w$ after which the plot of $\overline{D}_k$ vs. $w_k$ has approximately flattened out. This choice of RVE agrees quite well with the results from the FE simulation. 

We also note that the shape of the $\overline{D}_k$ vs. $w_k$ curves in Figs. \ref{fig:fig3}(b, c) are all quite similar with the elbow in the same location. This indicates that the approach is relatively robust to the particular ML model (as long as the model provides a good fit to the data per some appropriate CV measure) and to choices of $\bm{A}$. 

\subsection{Microstructure with irregularly shaped particles}\label{sec:irregular}
\begin{figure*}[ht!]
    \centering
    \includegraphics[width=\textwidth]{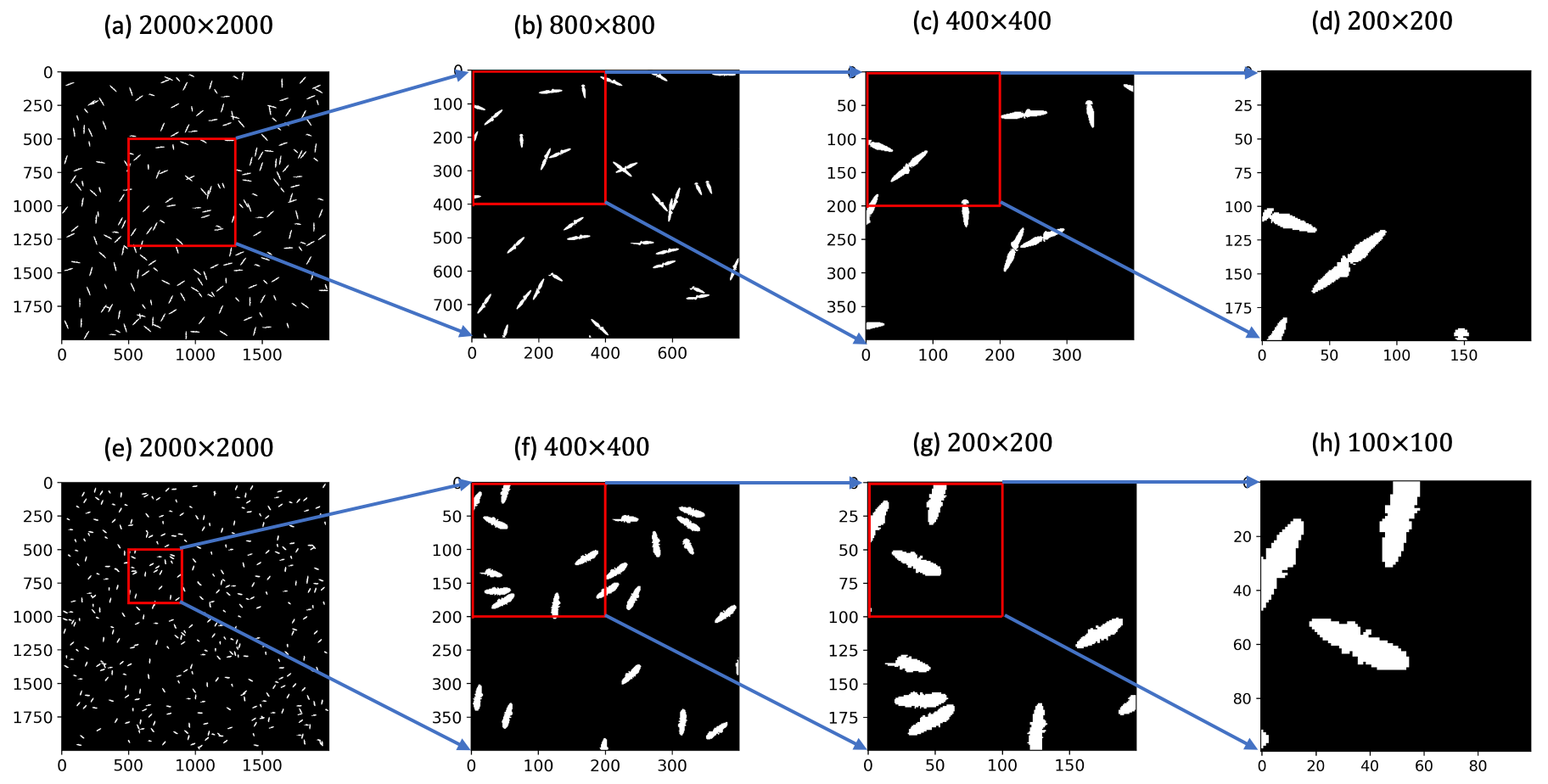}
    \caption{Micrographs with irregularly-shaped particles with the top and bottom rows representing higher- and lower-resolution images, respectively, of the same microstructure. The $2000\times2000$ pixel images in (a) and (e) represent $15 \mu m\times 15 \mu m$ and $30 \mu m\times 30\mu m$ physical regions. (b - d) and (f - h) are magnified versions of smaller windows from (a) and (e), respectively.}
    \label{fig:fig4}
\end{figure*}

We next consider micrographs with irregularly shaped particles, and we also investigate the effect of micrograph resolution on our selected RVE size. Regarding the latter, if one considers micrographs of the same microstructure but at different resolutions, the hope would be that our algorithm applied to either resolution would select the same RVE size in terms of its physical dimension (assuming both resolutions were fine enough to capture the shapes of the particles). To investigate this, we consider the two micrographs in Figs. \ref{fig:fig4}(a, e). The two micrographs are of the same microstructure, but Fig. \ref{fig:fig4}(a) is higher resolution with $2000$ pixels representing $15 \mu m$ physical length, compared to $2000$ pixels representing $30 \mu m$ physical length in Fig. \ref{fig:fig4}(e). Figs. \ref{fig:fig4}(b, f), \ref{fig:fig4}(c, g), and \ref{fig:fig4}(d, h) show magnified versions of variously located smaller windows within the original micrographs in Figs. \ref{fig:fig4}(a, e). The pair of images in each column represent the same physical size but at different resolutions. The particles in the bottom row appear thicker than in the top row because they have been magnified from the lower-resolution micrograph in Fig. \ref{fig:fig4}(e).

\begin{figure*}[ht!]
    \centering
    \includegraphics[width=0.85\textwidth]{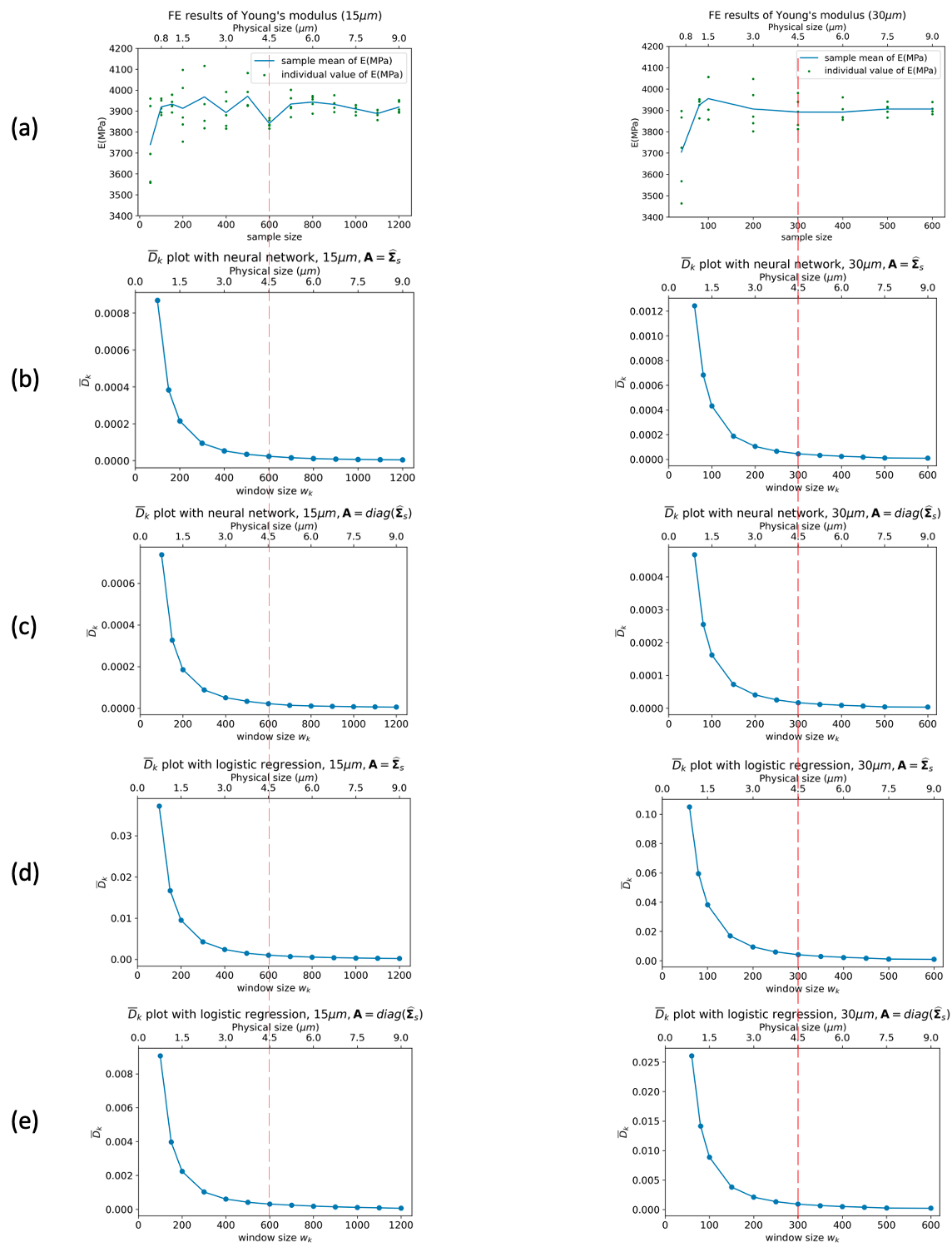}
    \caption{(a) Plots of FE-simulated Young's modulus vs. sample size for the micrographs in Fig. \ref{fig:fig4}(a) ($15\mu m\times 15\mu m$, left column) and Fig. \ref{fig:fig4}(b) ($30\mu m\times 30\mu m$, right column), where five samples were simulated for each sample size. Green dots are the simulated properties of individual samples, and the blue line is the mean of the five individual sample results. (b - e) Plots of $\overline{D}_k$ vs. $w_k$ for $15\mu m\times 15\mu m$ (left column) and $30\mu m\times 30\mu m$ (right column) micrographs. Plots in (b - c) were for neural networks, and plots in (d - e) were for logistic regression models. The red vertical lines indicate the ground truth RVE size based on the FE results in (a).}
    \label{fig:fig5}
\end{figure*}

Fig. \ref{fig:fig5} shows results for this example analogous to those shown in Fig. \ref{fig:fig3}. The FE results in Fig. \ref{fig:fig5}(a) indicate that RVE size is approximately $600$ pixels for the $15\mu m$ micrograph and $300$ pixels for the $30\mu m$ one, both of which correspond to $4.5\mu m$ physical length. We applied our RVE size selection algorithm with the micrographs in Figs. \ref{fig:fig4}(a, e) as the input micrographs. The CV accuracies of the fitted neural networks for the micrographs in Figs. \ref{fig:fig4}(a, e) are $98.31\%$ and $98.34\%$, respectively, and the CV accuracies of the fitted logistic regression models for Figs. \ref{fig:fig4}(a, e) are $98.04\%$ and $97.95\%$, respectively. Figs. \ref{fig:fig5}(b - e) show the plots of $\overline{D}_k$ vs. $w_k$ for our method applied to the higher (left column) and lower (right column) resolution micrographs, for the two different choices for $\bm{A}$ and for the two different ML models. The plots of $\overline{D}_k$ vs. $w_k$ stabilize at around $w=600$ pixels for the $15\mu m$ micrograph and $w=300$ pixels for the $30\mu m$ micrograph, which correspond to the same physical length of $4.5\mu m$ that the FE results indicate is the RVE size. This consistency when the $\overline{D}_k$ vs. $w_k$ plots stabilize indicates that our method is relatively robust to micrograph resolution, as long as they are high enough resolution to capture the actual shapes of the particles. Moreover, as is the case in Fig. \ref{fig:fig3}, the results in Fig. \ref{fig:fig5} indicate that our approach is relatively robust to choices of the particular ML model and the $\bm{A}$ matrix.

\subsection{Microstructure with agglomerations}\label{sec:agglo}
\begin{figure*}[ht!]
    \centering
    \includegraphics[width=\textwidth]{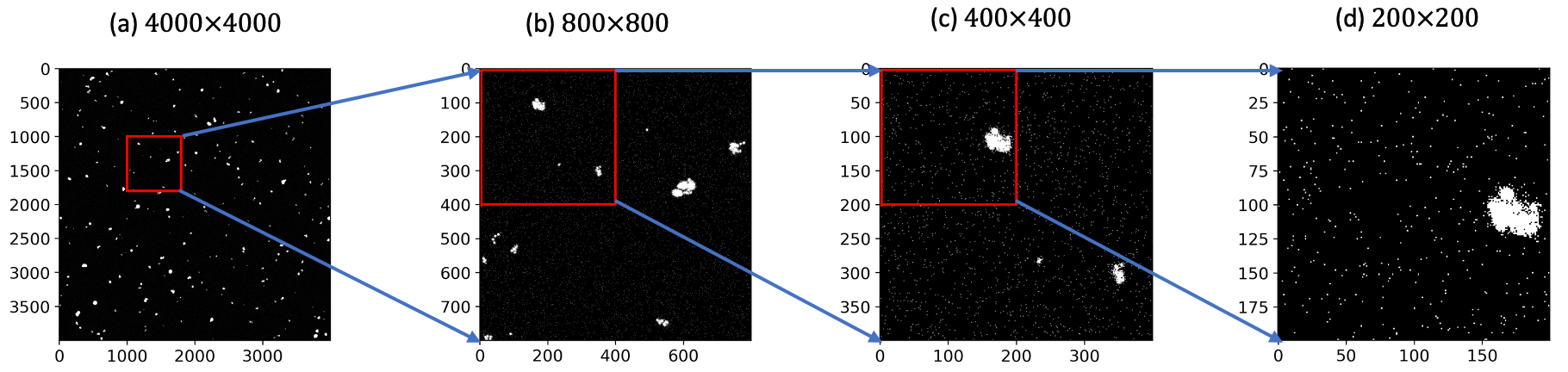}
    \caption{(a) A $4000 \times 4000$ pixel \text{micrograph} that represents a $60\mu m \times 60\mu m$ microstructure with $2\%$ vf. (b-d) Magnified subregions of (a) with various window sizes.}
    \label{fig:fig6}
\end{figure*}

We next consider a microstructure with agglomerations of the particles. Fig. \ref{fig:fig6}(a) shows a $4000\times 4000$ pixel micrograph representing a $60\mu m\times 60\mu m$ microstructure with $2\%$ vf. In Fig. \ref{fig:fig6}(a), the small bright dots are agglomerations of many particles, since the individual particles are too small to be discernible at this magnification level. In Fig. \ref{fig:fig6}(d), the individual particles become discernible, and the agglomerations appear as agglomerations. 

\begin{figure*}[ht!]
    \centering
    \includegraphics[width=\textwidth]{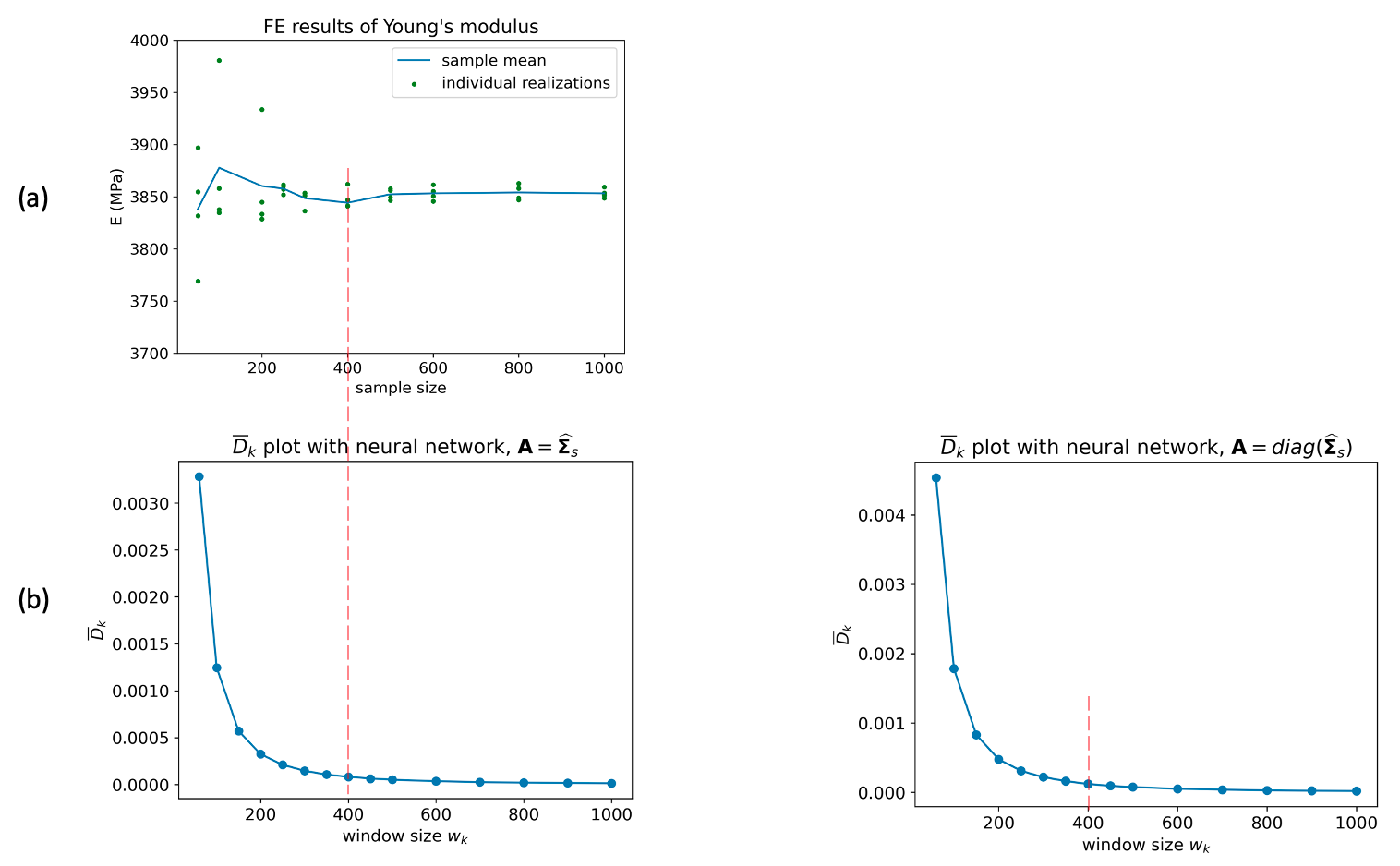}
    \caption{(a) FE-simulated Young's modulus for the micrograph in Fig. \ref{fig:fig6}(a) with four samples simulated for each sample size. Green dots are the simulated properties of individual samples, and the blue line is the mean of the four individual sample results. (b) Plots of $\overline{D}_k$ vs. $w_k$ for a neural network ML model. Left plot used $\bm{A} = \widehat{\bm{\Sigma}}_s$ and right plot used $\bm{A}= diag(\widehat{\bm{\Sigma}}_s)$. The red vertical lines indicate the ground truth RVE size based on the FE results in (a).}
    \label{fig:fig7}
\end{figure*}

Fig. \ref{fig:fig7}(a) shows the FE results for Young's modulus, from which we conclude that the ground truth RVE size is approximately $400$ pixels. Figs. \ref{fig:fig7}(b, c) show the plots of $\overline{D}_k$ vs. $w_k$ from our RVE selection algorithm for a neural network ML model (CV accuracy $71.86\%$) for the two choices for $\bm{A}$. Note that the lower CV accuracy of the model compared to the results of the previous two examples is due to the inherent nature of this classification problem (it is not possible to nearly perfectly classify each pixel value). Our nonstationarity monitoring and RVE determination algorithm appears to still work well even when the predictive accuracy of the supervised learning model is lower, as long as the model is sufficient to represent a stochastic fingerprint of the microstructure, in the sense that the nonstationarity of the microstructure is reflected in the Fischer score vectors. A similar phenomenon was also observed in the prior work on the score-based nonstationarity monitoring\cite{Kungang1, ZHANG2021116818}. In particular, for examples in which both a neural network and a logistic regression model were fit and the neural network achieved better predictive accuracy, the nonstationarity monitoring based on the less accurate logistic regression model was still almost as effective. We did not include a logistic regression model in this example, since the previous examples indicated that the two ML models give comparable results. The $\overline{D}_k$ vs. $w_k$ plots both stabilize at around $w = 400$ pixels, which again agrees with the ground truth RVE size from Fig. \ref{fig:fig7}(a). 

As we discussed earlier, our approach is intended to give an upper bound on the RVE size. However, in all three examples that we considered, the suggested RVE size appears to be a relatively tight upper bound, since it agrees quite closely with the convergence of the FE-simulated properties. 

\section{Conclusions}\label{sec: conclusion}
In this paper, we have presented a new approach for RVE size determination. The primary advantage of our approach is that it determines RVE size based only on a sufficiently large micrograph of the microstructure, with no need for FE simulation. In contrast, existing methods require many FE simulations of multiple samples of each size, for sample sizes varying from small samples to samples that are larger than an RVE.

The approach involves fitting any appropriate standard ML model to implicitly characterize the stochastic nature of the microstructure, computing the Fisher score vectors for the model, and then determining the smallest window size over which the average score vector ($\overline{D}_k$ for window size $w_k$) varies negligibly as the window moves across the micrograph. The rationale behind the approach is well-rooted in fundamental statistical theory regarding Fisher score vectors. To validate our approach, we compared the RVE size that it selects with the "ground truth" RVE size based on extensive FE simulation to determine when the FE-simulated properties actually converge. In all of the examples that we considered, our selected RVE size agreed quite well with the ground truth RVE size.

Throughout the paper we have illustrated the approach using two-phase microstructures for which linear elastic FE models are appropriate. However, because our RVE size determination is based entirely on the microstructure characteristics and not on the physics of the FE simulation, it is also applicable to more complex FE simulations, e.g., of nonlinear viscoelastic behavior\cite{QIAO2009491}, providing the properties of interest are determined solely by the morphology (micrograph) of the material. One example of exception could be the study of properties of two-phase nanocomposites consisting of nanoparticles dispersed in a polymer matrix, where the interphase characteristics may depend on the properties of both filler and the matrix as well as morphology, so that two micrographs that have different fillers and matrices but are otherwise identical may have different properties and different RVE sizes\cite{Ying12,HU201855,Xu2021}. In such situations, we anticipate that our approach would still be applicable if the micrographs were augmented with information on the interphase behavior (i.e., the same interphase information that must be specified to conduct an FE simulation of the material. Potentially, this could be achieved by converting the micrograph from two-phase to three-phase, where the third phase represents the interphase. Another potential future direction is extending this approach to polycrystalline microstructures. In theory, as long as one can choose an appropriate supervised learning model that can implicitly characterize the stochastic nature of the microstructure (in the sense that the average score vector is indicative of microstructure nonstationarity), we expect our approach to be applicable for polycrystalline microstructures. However, the binary microstructure images that we have used for the two-phase composites would have to be replaced by images showing grain boundaries annotated with information on crystal system and orientation within each region. Finally, we have developed our approach around 2D material samples. Extension to 3D samples and 3D simulation is conceptually trivial, with the square 2D moving windows replaced by 3D moving boxes. The primary challenge would be computational expense, but this is also the case for extending most mechanics methods from 2D to 3D. 

\begin{acknowledgments}
This work was funded by the Air Force Office of Scientific Research Grant $\#$FA9550-18-1-0381, which we gratefully acknowledge. We appreciate Prof. Anton Van Beek, who provided the reconstruction algorithm that we used to generate all micrographs. This work made use of the Quest high-performance computing facility at Northwestern University, which is jointly supported by the Office of the Provost, the Office of Research, and Northwestern University Information Technology.
\end{acknowledgments}

\section*{Author Declarations}
\subsection*{Conflict of Interest}
The authors have no conflicts to disclose.

\subsection*{Author Contributions}
\textbf{Wei Liu}: Conceptualization (equal); Formal analysis (equal); Investigation (equal); Methodology (equal); Software (equal); Validation (equal); Visualization (equal); Writing – original draft (equal). \textbf{Satyajit Mojumder}: Data curation (equal); Methodology (equal); Resources (supporting); Validation (equal); Writing – review \& editing (equal). \textbf{Wing Kam Liu}: Funding acquisition (equal); Project administration (equal); Supervision (equal); Writing – review \& editing (equal). \textbf{Wei Chen}: Funding acquisition (equal); Project administration (equal); Supervision (equal); Writing – review \& editing (equal). \textbf{Daniel W. Apley}: Conceptualization (equal); Funding acquisition (equal); Methodology (equal); Project administration (equal); Resources (equal); Supervision (equal); Writing – review \& editing (equal).

\section*{Data Availability Statement}
The data that support the findings of this study are openly available in Github at \url{https://github.com/EurusWei/RVE}.

\nocite{*}
\bibliography{RVE}

\end{document}